\documentclass[twocolumn,prl,showpacs,float]{revtex4}  

\usepackage{graphicx}
\usepackage{float}

\usepackage{epsfig}

\begin{document}

\title { A hexatic smectic phase with algebraically decaying
  bond-orientational order
  }

\author{Lorenzo Agosta$^{1}$ Alfredo Metere$^{2*}$ and  Mikhail Dzugutov$^3$}

\affiliation{$^1$ Department of Materials and Environmental Chemistry,
  Stockholm University, Arrhenius V\"{a}g. 16C S-10691 Stockholm,
  Sweden\\$^2*$ Physical and Life Science Directorate, Computational
  Materials Science in the Condensed Matter and Materials Division,
  Lawrence Livermore National Laboratory, 7000 East Avenue L-367,
  Livermore, CA - 94550, USA\\$^3$ Department of Mathematics, Royal
  Institute of Technology, SE-100 44 Stockholm, Sweden\\ }

\begin{abstract}
  The hexatic phase predicted by the theories of two-dimensional
  melting is characterised by the power law decay of the orientational
  correlations whereas the in-layer bond orientational order in all
  the hexatic smectic phases observed so far was found to be
  long-range. We report a hexatic smectic phase where the
  in-layer bond orientational correlations decay as $\propto
  r^{-1/4}$, in quantitative agreement with the hexatic ordering
  predicted by the theory for two dimensions. The phase was formed in
  a molecular dynamics simulation of a one-component system of
  particles interacting via a spherically symmetric potential. This is
  the first observation of the theoretically predicted two-dimensional
  hexatic order in a three-dimensional system.

  \end{abstract}

\date{\today}


\maketitle

The theory of two-dimensional (2D) melting by Kosterlitz, Thouless,
Halperin, Nelson and Young (KTHNY)\cite{Kos} predicts the existence of
a distinct new phase intervening between a solid and a liquid. This
phase, called hexatic, is a 2D fluid characterised by a quasi-long
range bond orientational order (BOO) (decaying as power law) and
short-range (exponentially decaying) positional correlations. The
hexatic phase predicted by the KTHNY theory has been observed in a
number of real 2D systems \cite{CLARK}, but the attempts to find it in
three-dimensional (3D) systems have so far been
unsuccessful. Nevertheless, its terminology has been carried over to
3D liquid crystals \cite{GENNES} to describe the bond-ordered liquid
states found in the axially stacked layers of some smectic liquid
crystals \cite{Jeu, GOOD, ZALU}. These smectic phases, calles hexatic
smectics, were thus suggested to be the 3D analog of the 2D hexatic
phase conjectured by the KTHNY scenario \cite{BIRG}. It has to be
stressed, however, that this analogy is purely heuristic. The
principal difference between the two phases is that the hexatic
smectics exhibit true long-range in-layer BOO in contrast to its power
law decay in the 2D hexatic phases.  This difference was tentatively
attributed to the interaction between the smectic layers and the
effect of anisotropic forces \cite{BIRG}, but the nature, and the
origin of the long-range BOO in the hexatic smectic phases still elude
comprehensive understanding.

Particle simulations have been actively used to understand the
formation mechanism of the smectic liquid crystals in terms of the
molecular-level properties \cite{WILSON}. Following the seminal work
of Onsager \cite{ONS}, it was commonly believed that formation of
smectic phases is driven by the packing entropy of anisometric
(rod-like) mesogenic molecules \cite{LUB}. Accordingly, a rod-like
particle shape was assumed in the computer models of smectic phases
\cite{FRE, MIGUEL}. However, no unconstrained simulation of a
hexatic smectic phase has so far been reported \cite{MARTINEZ}.

Two questions of general conceptual interest arise in this
context. (i) Is the anisometry of the mesogenic molecules a
prerequisite for producing a smectic mesophase and, in particular, a
hexatic smectic phase? (ii) Can the true long-range BOO observed in
the hexatic smectic phases be related to the specific shape of their
constituent molecules and the anisotropy of the intermolecular forces?

In this Letter, we report a molecular-dynamics simulation addressing
these questions. It is demonstrated that a single-component system of
particles interacting via a spherically-symmetric potential forms an
equilibrium hexatic smectic mesophase where the in-layer BOO decays as
a power law, in  quantitative agreement with the KTHNY theory
prediction.

\begin{figure}[h] 

\includegraphics[width=5.5cm]{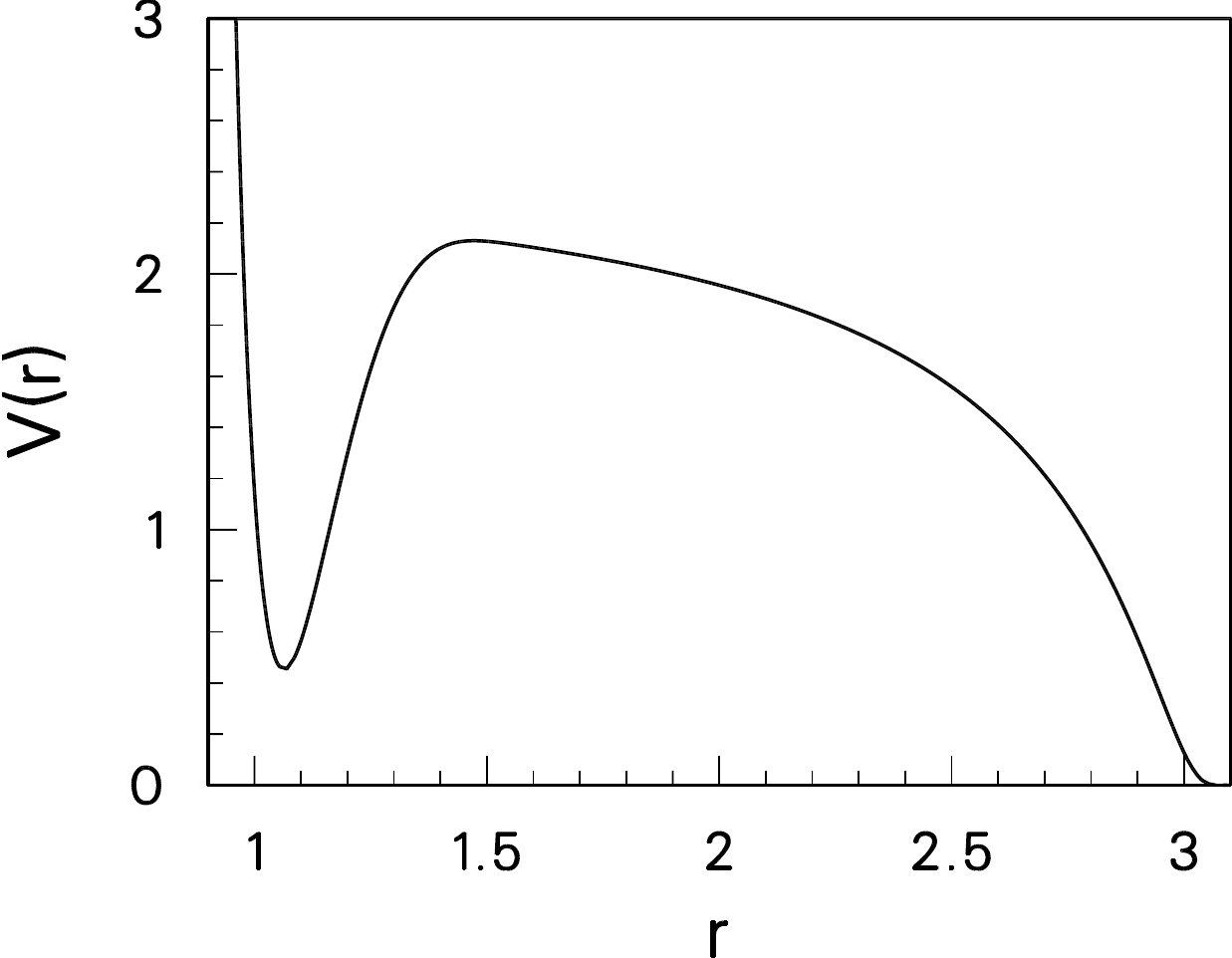}
\caption{ Pair potential}
\label{fig1}
\end{figure}

We investigated a molecular-dynamics model of 50000 identical
particles confined to a cubic box with periodic boundary conditions
interacting via the pair potential shown in Fig.\ref{fig1}. The
functional form of the potential energy for two particles separated by
the distance $r$ is:
\begin{equation} 
V(r)  = a_1  ( r^{-m} - d) H(r,b_1,c_1) + a_2 H(r,b_2,c_2)   
\end{equation}
\begin{equation} 
H(r,b,c) = \left\{ \begin{array}{ll}
  \exp\left( \frac{b}{r-c} \right) & r < c\\
0 & r \geq c
\end{array}
\right.       
\end{equation}
\begin{table} [h]
\begin{tabular}{cccccccc}
 \hline 
\hline 

m & $a_1$ & $b_1$ & $c_1$ & $a_2$ & $b_2$ & $c_2$ & $d$ \\

\hline 
12 \space & 113 \space & 2.8 \space & 1.75 \space & 2.57 \space & 0.3 \space &
3.1 \space & 1.4\\
\hline 
\end{tabular}
\caption{Values of the  parameters for the pair potential used in this
  simulation (Eq 1,Fig.\ref{fig1}).} 
\label{table1}
\end{table}
The values of the parameters are presented in Table \ref{table1}.  The
simulation reduced units are those used in the definition of the
potential. This pair potential represents a modification of an earlier
reported one \cite{DZ1} that was found to produce a smectic-$B$
crystal. The main difference between the two potentials is that in the
present one the long-range repulsion is extended to a significantly
larger distance. In that earlier simulation the latter parameter was
found to determine the interlayer spacing.

\begin{figure} 
\includegraphics[width=6.cm]{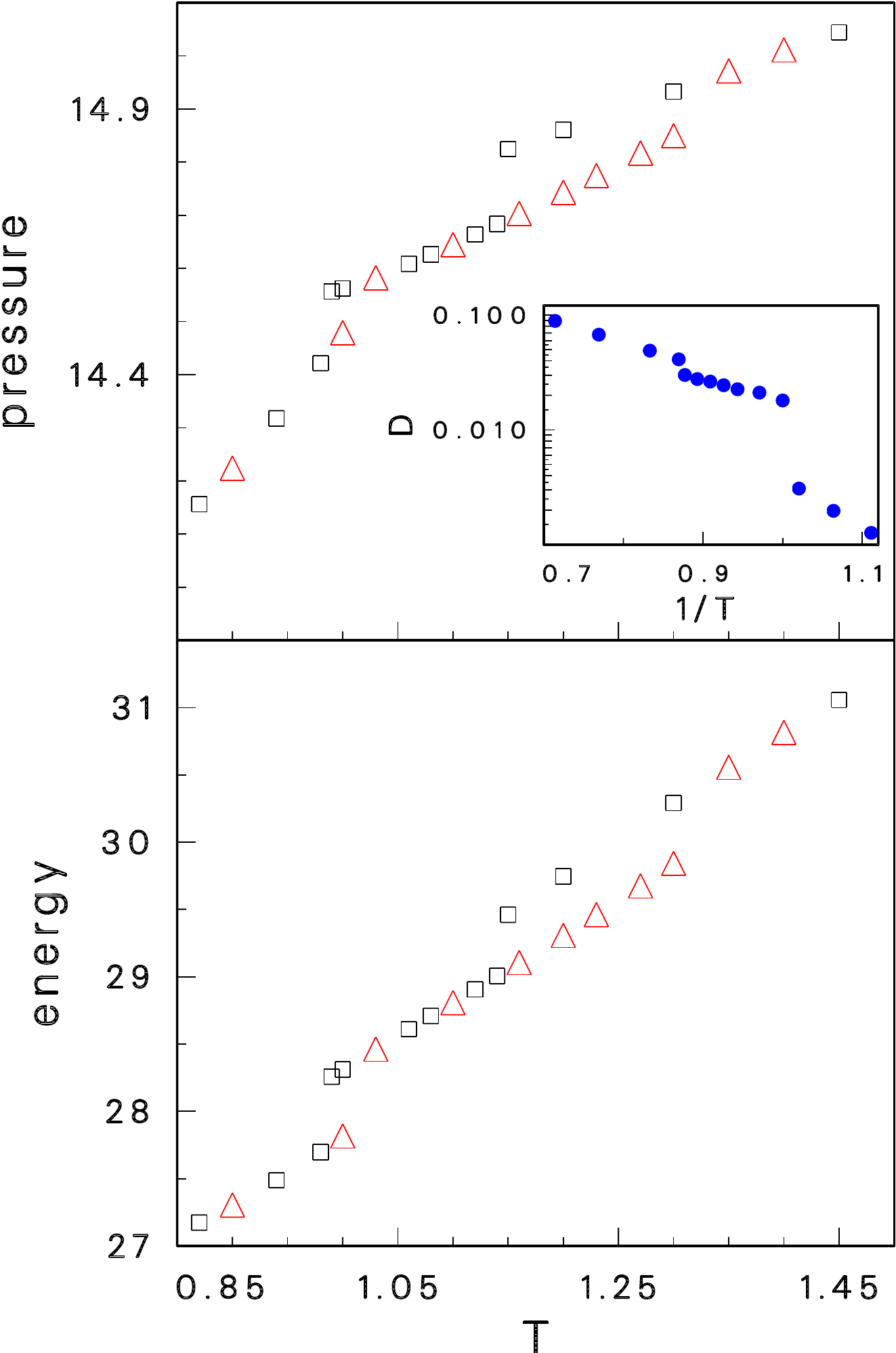}
\caption{ Temperature variation of the pressure and energy at the
  number density $\rho = 0.41$. Squares: cooling; triangles:
  heating. Inset: the Arrhenius plot of the diffusion
  coefficient.}
\label{fig2}
\end{figure}

The system's phase behaviour was investigated at a constant number
density $\rho=0.41$.  The temperature was changed in a stepwise
manner, performing a comprehensive equilibration after each step which
typically amounted to $10^7$ timesteps. The simulation started by
equilibrating an isotropic liquid state at sufficiently high
temperature.  Fig. \ref{fig2} shows system's energy and pressure as
functions of temperature. Upon cooling, both quantities exhibited a
discontinuity at $T=1.15$, followed by another one at $T=0.95$. The
latter was accompanied by a sharp drop in the diffusion rate,
Fig. \ref{fig2}, indicating the formation a solid state; this was
identified as a smectic B crystal \cite{suppl}. Upon re-heating the
described temperature variations of the pressure and energy were
reproduced. Each observed singularity was found to be accompanied by a
hysteresis, a signature of the first-order nature of the respective
transition.

\begin{figure}  
\includegraphics[width=6.5cm]{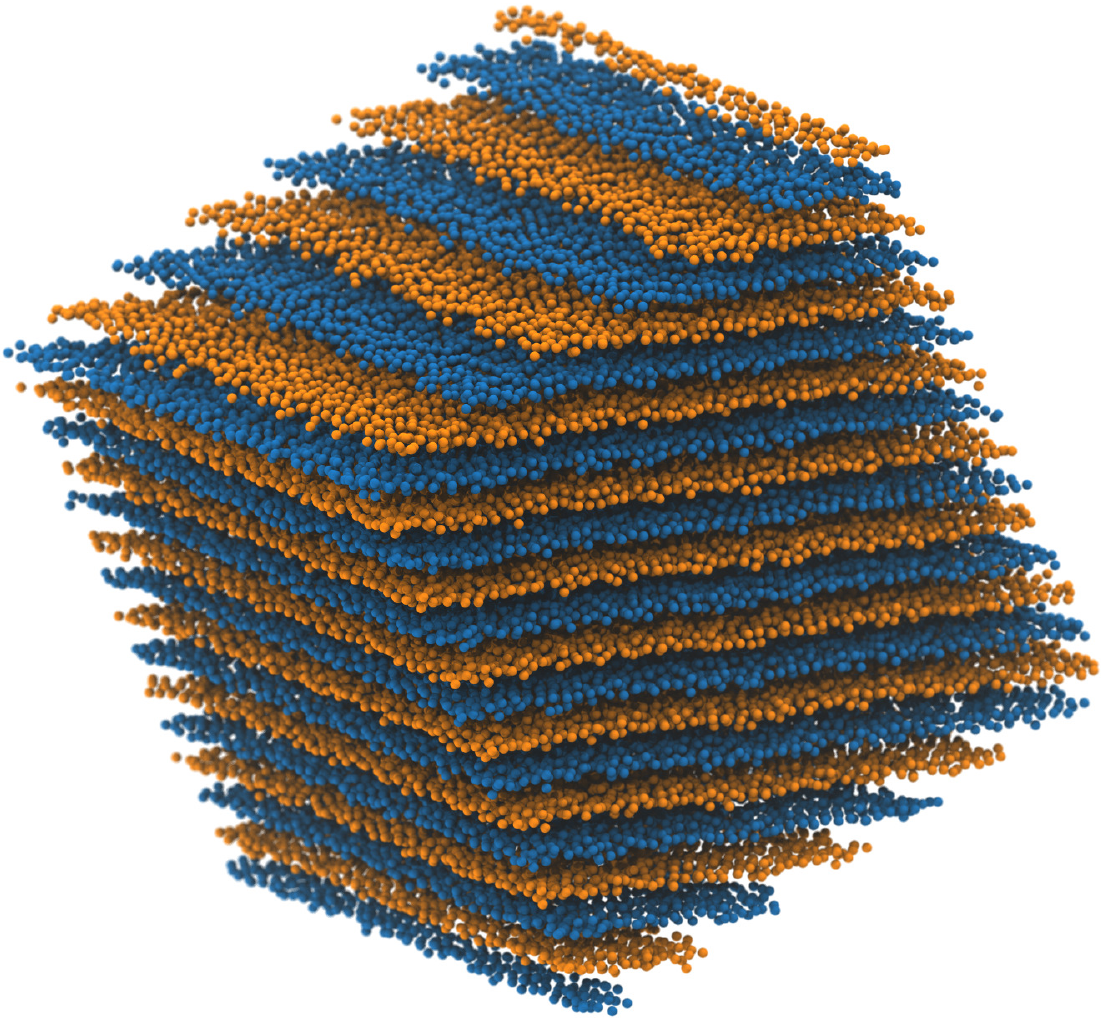}
\caption{ A view of the simulated smectic phase, adjacent
  layers are discriminated by color}
\label{fig3}
\end{figure}

The observed phase behaviour thus demonstrates the existence of a
distinct equilibrium fluid phase interposed between the isotropic
liquid and the Smectic $B$ crystal, separated from each of the latter
two phases by a first-order transition. The general view of its
instantaneous configuration presented in Fig. \ref{fig3} suggests
that this is a smectic liquid crystal composed of uniaxially stacked
layers with a liquid-like in-layer diffusion, Fig. \ref{fig2}.
We note that its estimated interlayer spacing \cite{suppl} is
consistent with the long-range repulsion distance of the pair
potential, Fig.\ref{fig1}.

\begin{figure} [h] 
\includegraphics[width=9.cm]{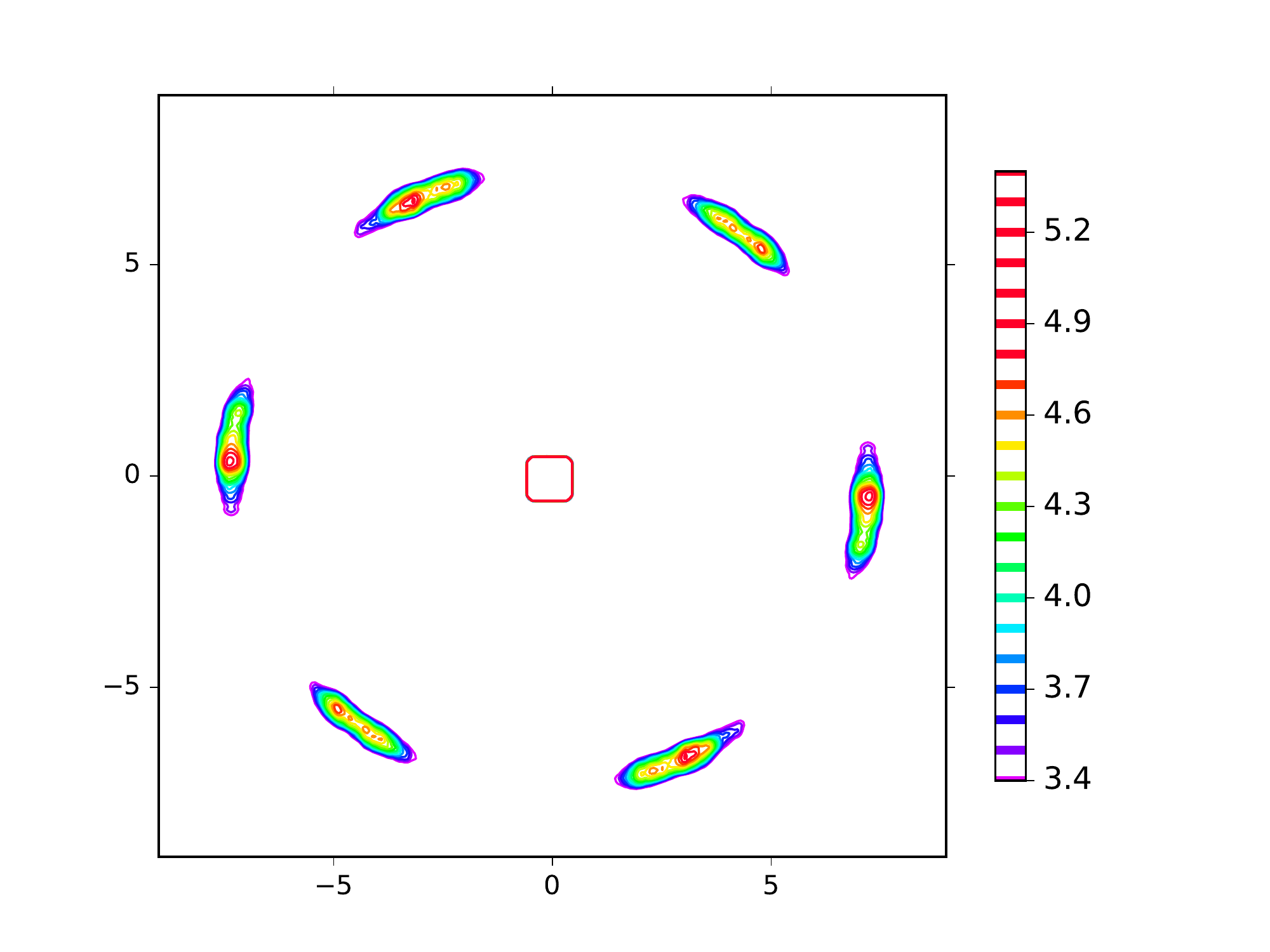}
\includegraphics[width=5.cm]{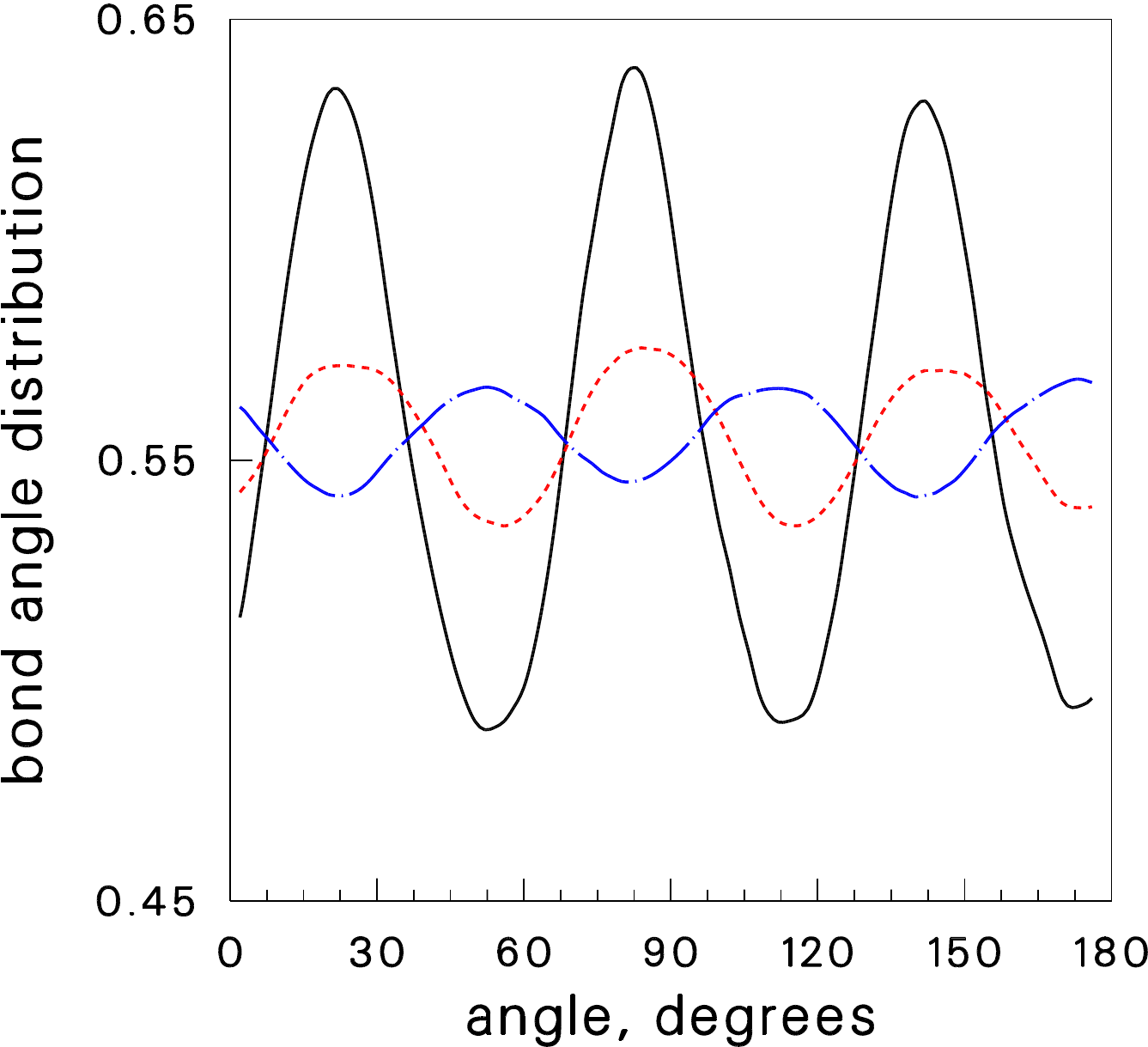}
\caption{(a) $S({\bf Q})$ of a single layer calculated in the layer
  plane. (b) Bond angle distribution.  Solid line and dash-dotted line,
  respectively: one layer at $T=1.0$ and at $T=1.1$. Dashed line:
  entire system, $T=1.0$. }
\label{fig4}
\end{figure}
\begin{figure}   
\includegraphics[width=7.cm]{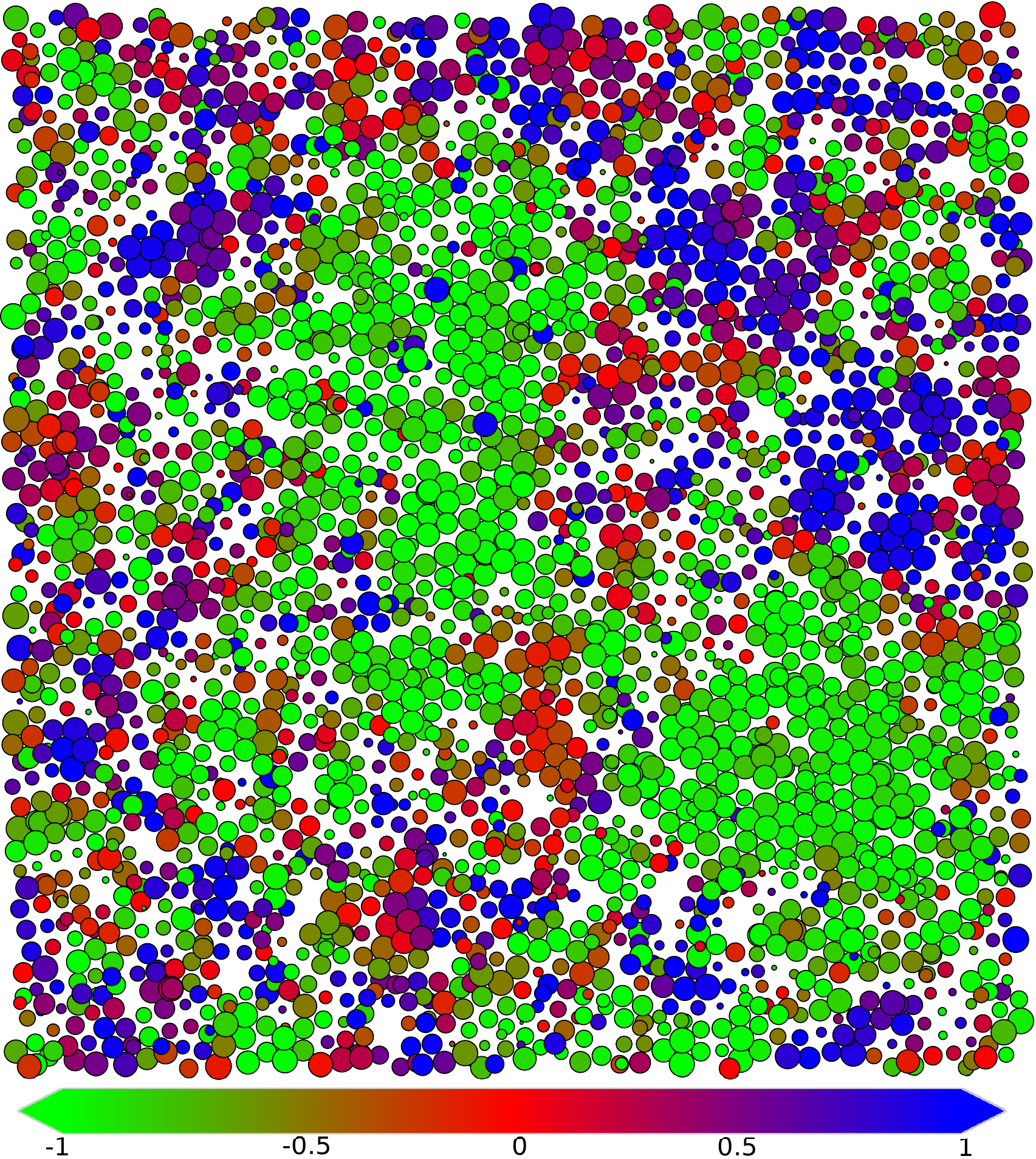}
\par
\vspace{0.5cm}
\includegraphics[width=7.5cm]{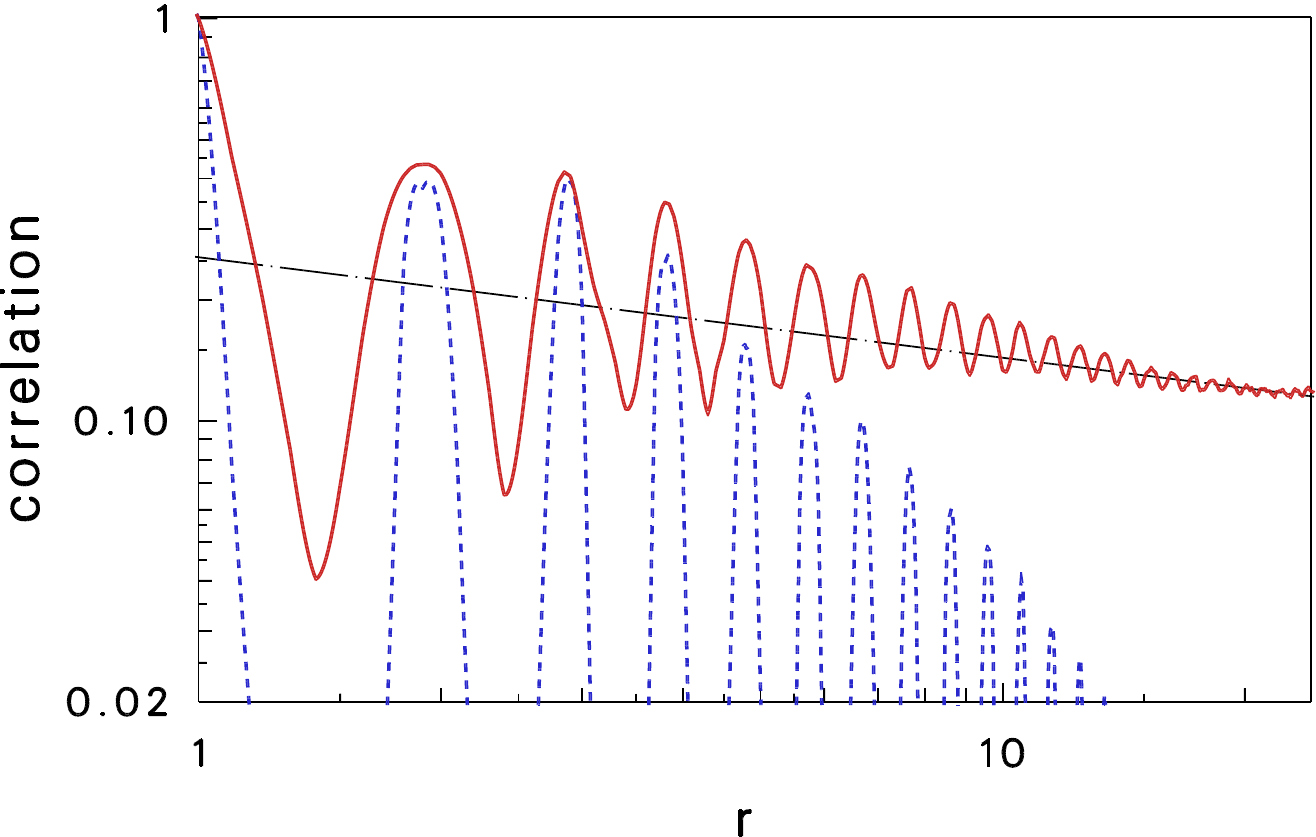}
\caption{ (a) The local BOO distribution in a layer at $T=1$. Each dot
  represents a particle; the size of the dot representing particle $j$
  is proportional to $|\Psi({\bf r}_j)|$, its color indicates
  $Re(\Psi({\bf r}_j) )$ according to the scale. (b) Solid line: BOO
  correlation function $g_6(r)$; dashed line: $g(r)-1$, both at $T=1.$;
  dash-dotted line: $ \propto r^{-1/4}$.}
\label{fig5}
\end{figure}
In order to understand the exact nature of thus produced smectic
mesophase we performed a detailed analysis of its in-layer
structure. As a first step in the structure characterisation we
calculated the structure factor $S({\bf Q})$ \cite{suppl} representing
the scattered intensity in the diffraction experiments. Having
established the global uniaxial symmetry of the configuration and the
axis orientation \cite{suppl}, we then calculated $S({\bf Q})$ in the
layer plane $Q_z=0$, $Q_z$ being the axis coordinate.  Fig.\ref{fig4}a
shows $S({\bf Q})$ for a single layer averaged over $10^4$ timesteps.
It exhibits a pronounced azimuthal modulation in the form of six
diffuse arcs characteristic of the diffraction patterns of hexatic
smectic phases. Their radial position can be identified as $Q=4\pi /
(a \sqrt 3)$ where $a$ is the in-layer nearest neighbour distance. We
notice that this distance is in good agreement with the position of
the first potential minimum, Fig. \ref{fig1}. In this way, the short
repulsion and the long repulsion parts of the pair potential act,
respectively, as the length and the diameter of the mesogenic
molecules forming the real smectic phases: the former define the
interlayer particle packing whereas the latter define the interlayer
spacing.

The sixfold angular symmetry of the diffraction pattern is a necessary
but not sufficient condition for identifying the simulated phase as a
hexatic smectic.  To get further evidence for the hexatic nature of
its in-layer structure we calculated the bond-orientation distribution
which is shown in Fig.  \ref{fig4}b.  The bonds were defined as the
pairs of particles within a layer separated by the nearest-neighbour
distance $a$ as indicated above. The angles presented in the
statistics were measured between the bonds and an axis chosen in the
layer plane. The statistics was calculated for an ensemble of
configurations produced within a simulation run of $10^4$ time
steps. The bond angle distribution for a single layer at $T=1.0$
demonstrates a pronounced six-fold modulation with the amplitude
consistent to that observed in the azimuthal variation of $S({\bf
  Q})$, Fig.  \ref{fig4}a. The amplitude of the distribution
modulation for the same layer at $T=1.1$ is significantly smaller, as
well as the one calculated for the entire system.

Next, we analyse the pattern of the local six-fold BOO in a layer
configuration.  For each particle position ${\bf r}_j$ we calculated a
vector $\Psi({\bf r}_j)= \frac{1}{N_k} \sum_{k=1}^{N_k}\mathrm{e}^{i 6
  \theta_{jk}}$ where $ \theta_{jk}$ is the angle formed by the bond
linking particle $j$ with its nearest neighbour $k$ relative to an
arbitrary axis, and $N_k$ is the number of the nearest neighbours. Fig
\ref{fig5}a shows the distribution of these vectors in a layer at
$T=1.0$. Each vector $\Psi({\bf r_j})$ is represented by a dot; the dot's
size is proportional to $|\Psi({\bf r_j})|$ and the vector orientation
is indicated by the dot's colour, according to the scale. The
distribution exhibits an apparent domain structure. A cluster of
coherent hexagonal order percolates through the entire layer, which
can account for the six-fold symmetry breaking in both in the
diffraction pattern and in the bond-angle distribution. Besides, there
are twinning domains of hexagonal order rotated by $30^{\circ}$ and
$15^{\circ}$ with respect to the main domain. These domains can be
discerned in the pattern of bonds produced for the same particle
configuration \cite{suppl}.

The identifying feature of the hexatic phase according to the KTHNY
theory is the algebraic decay of its BOO.  The latter can be
quantified as follows:
\begin{equation} 
  g_6(r) =  \frac{ \langle   \sum_{k \neq j}^N  \Psi({\bf r}_j)
    \Psi({\bf r}_k) \delta ( r - |{\bf r}_j - {\bf r}_k| )  \rangle} 
{\langle   \sum_{k \neq j}^N  \delta ( r - |{\bf r}_j - {\bf r}_k|)  \rangle }
\end{equation}
where $N$ is the number of particles, and $\langle \rangle$ denote
ensemble averaging. Fig.  \ref{fig5}b shows $g_6(r)$ calculated for an
ensemble of configurations of a single layer produced in a simulation
run of $10^4$ time steps. It is compared with the radial distribution
function $g(r)$ \cite{HANSEN} expressing the decay of the positional
correlation. We find that the calculated $g_6(r)$ asymptotically
decays as $\propto r^\eta$ with $\eta=-1/4$ which is in  quantitative
agreement with the prediction of the KTHNY theory for the 2D hexatic
\cite{Kos}, whereas $g(r)$ decays exponentially. These results
explicitly prove that the layers of the simulated smectic represent 2D
hexatic phases as defined by the theory.

Three conceptually new aspects of this study deserve to be remarked.

First, the finding that a system of identical particles interacting
via a spherically-symmetric potential can form a hexatic smectic phase
changes the basic model of smectic phases, thereby advancing our
understanding of the causes underlying the occurrence of particular
structures in the phase transformations of liquid crystals.

Second, the observed algebraic power-law decay of the in-layer BOO in
a hexatic smectic phase formed by a system of particles with
spherically-symmetric interaction suggests that the true long-range
BOO that has so far been found in the hexatic smectics can be
attributed to the rod-like shape of their constituent molecules and
the anisotropy of the intermolecular forces.

Third, the hexatic phase predicted by the KHTNY theory of 2D melting
has so far never been found in a 3D system.  The smectic phase we
report here demonstrates the in-layer hexatic order that
quantitatively agrees with the theory's prediction. This is the first
indication that the theory's application scope can include 3D systems.

We note that the pair potential we report is similar to that predicted
for colloidal systems \cite{VO} (amended with steric repulsion)
suggesting that a hexatic smectic phase can be formed by spherical
colloidal particles with an appropriately tailored interaction, as
microgels or through a cosolute \cite{Zac}.

In summary, we report a hexatic smectic phase formed in a
molecular dynamics simulation of a one-component system of particles
interacting via a spherically symmetric potential. In contrast to the
hexatic smectics observed so far, its BOO decays algebraically in 
quantitative agreement with the KTHNY theory prediction for the 2D
hexatic phase.  This is the first hexatic smectic phase produced in a
particle simulation, and the first observation of the 2D hexatic phase
in a 3D system.
\\

\section{Acknowledgements}
We thank Dr. Babak Sadigh for his illuminating suggestions. Simulations were performed using GROMACS. This work was performed under the auspices of the U.S. Department of Energy by Lawrence Livermore National Laboratory under Contract XX-XXXX-XXXXXXXXX.


\end{document}